\documentclass{article}
\usepackage{geometry}
\usepackage{graphicx, color, xurl, amssymb, dsfont, subcaption}
\usepackage{xcolor}
\usepackage{makecell} 
\usepackage{colortbl}
\usepackage{glossaries}
\usepackage{svg}
\usepackage{amsmath}
\usepackage{algpseudocode} 
\usepackage{multirow}
\usepackage{booktabs}
\usepackage{enumitem}  
\usepackage{footnote}
\makesavenoteenv{tabular}
\makesavenoteenv{table}
\usepackage{amsthm, eurosym}
\usepackage{subcaption}
\usepackage{tabularx}
\usepackage[round, authoryear]{natbib}
\usepackage[linesnumbered]{algorithm2e}
\usepackage{nicefrac}
\usepackage{pgfplots}
\usepackage{hyperref}
\usepackage{graphicx}

\newcommand{\Dc}{\ensuremath{\mathcal{D}}}
\newcommand{\Fc}{\ensuremath{\mathcal{F}}}

\newcommand{\Hc}{\ensuremath{\mathcal{H}}}

\newcommand{\Mc}{\ensuremath{\mathcal{M}}}

\newcommand{\Oc}{\ensuremath{\mathcal{O}}}

\newcommand{\Sc}{\ensuremath{\mathcal{S}}}
\newcommand{\Tc}{\ensuremath{\mathcal{T}}}

\newcommand{\Xc}{\ensuremath{\mathcal{X}}}
\newcommand{\Ed}{\ensuremath{\mathds{E}}}

\newcommand{\Rd}{\ensuremath{\mathds{R}}}

\newcommand{\set}[1]{ \ensuremath{\left\{ #1 \right\}} }

\DeclareMathOperator*{\argmax}{arg\,max}

\title{Joint Bidding on Intraday and Frequency Containment Reserve Markets}
\author{Yiming Zhang\thanks{The Hong Kong Polytechnic University, Hong Kong; this work was mainly carried out at the Technical University of Munich and Entrix GmbH, Munich, Germany,  \url{yiming1.zhang@connect.polyu.hk}} \and Wolfgang Ridinger\thanks{Entrix GmbH, Munich, Germany, \url{wolfgang.ridinger@entrixenergy.com}} \and David Wozabal\thanks{Vrije Universiteit Amsterdam, Amsterdam, Netherlands, \url{d.wozabal@vu.nl}}}
\date{\today}

\definecolor{davidColor}{rgb}{0,0.7,0}
\definecolor{yimingColor}{rgb}{0,0,0.7}
\definecolor{wolfgangColor}{rgb}{0.7,0,0}
\definecolor{lightblue}{rgb}{0.3, 0.61, 0.98} 
\definecolor{lightgreen}{rgb}{0.56, 0.93, 0.56}

\begin{document}

\maketitle

\begin{abstract}
    As renewable energy integration increases supply variability, battery energy storage systems (BESS) present a viable solution for balancing supply and demand. This paper proposes a novel approach for optimizing battery BESS participation in multiple electricity markets. We develop a joint bidding strategy that combines participation in the primary frequency reserve market with continuous trading in the intraday market, addressing a gap in the extant literature which typically considers these markets in isolation or simplifies the continuous nature of intraday trading. Our approach utilizes a mixed integer linear programming implementation of the rolling intrinsic algorithm for intraday decisions and state of charge recovery, alongside a learned classifier strategy (LCS) that determines optimal capacity allocation between markets. A comprehensive out-of-sample backtest over more than one year of historical German market data validates our approach: The LCS increases overall profits by over 4\% compared to the best-performing static strategy and by more than 3\% over a naive dynamic benchmark. Crucially, our method closes the gap to a theoretical perfect foresight strategy to just 4\%, demonstrating the effectiveness of dynamic, learning-based allocation in a complex, multi-market environment.
\end{abstract}

\section{Introduction}
The increasing share of variable renewable sources of electricity (VRES), such as wind and solar energy, changes electricity systems around the world. As a consequence, unexpected and costly imbalances caused by the unpredictability of VRES output may lead to grid instability and even blackouts. Electricity storage can offer flexibility to help mitigate these problems and stabilize electricity systems, thereby generating welfare gains \citep{sioshansi2010welfare}. 

Traditionally, electricity storage is a scarce resource and pumped-hydro storage facilities have been the primary form of large-scale electricity storage available. These systems offer immense energy capacity, but their deployment is limited by strict geographical requirements~--~namely, the need for significant elevation differences between two large reservoirs. Consequently, the potential for expanding the capacity of pumped hydro storage is severely limited in most electricity systems, creating a pressing need for alternative flexibility solutions to accommodate the growing share of VRES.

To fill this gap, a new portfolio of flexibility technologies is emerging, including battery energy storage systems (BESS), green hydrogen for long-duration and seasonal energy storage, and demand response programs that provide flexibility by actively managing consumption patterns. Among these technologies, BESS stands out for their rapid response times and locational flexibility, making them particularly suitable for short-term energy arbitrage and ancillary services. Their high power-to-capacity ratio makes them well-suited for grid stabilization services, where supply-demand imbalances must be corrected on a second-to-second basis, and for participation in short-term electricity markets.

Due to the dramatic and ongoing reduction in battery costs, BESS are becoming an increasingly viable option for grid-level electricity storage. For instance, lithium-ion battery pack prices have seen a drop of more than $75$\% in the past $10$ years with prices decreasing from \$$463$/kWh in $2015$ to \$$115$/kWh in $2024$ \citep{bnef2024}. This downward trend is projected to continue, with costs expected to drop below \$100/kWh well before $2030$ \citep{nrel2023cost}.

In this paper, we propose a joint model for the participation in the market for primary frequency reserve and the intraday market (IDM), explicitly modeling orderbook-based continuous trading to maximize the potential to reoptimize positions multiple times. 

While a large number of literature addresses optimal bidding strategies for electricity storage, studies that jointly consider the intraday and frequency control markets remain rare. Most existing approaches, such as \citep{de2025,fk08}, focus exclusively on the day-ahead market. Although some papers incorporate both the day-ahead and intraday markets, the intraday market is typically represented as a single rebalancing decision. This simplification allows the decision maker to adjust positions taken in the day-ahead market but overlooks the continuous nature of intraday trading and the resulting opportunities for multiple rebalancing actions.
Examples of such papers include \citep{ff11,lwm13, kokf18, lohndorf2023}. \cite{RaWo18} consider the problem of coordinated bidding in sequential auctions for a renewable power producer without storage in the Spanish intraday market.

In recent years, the intraday market has attracted an increasing number of participants seeking to correct short-term deviations from planned generation and load schedules. Various trading strategies in the IDM were proposed. Dynamic programming approaches that model the intraday bidding process are proposed in \citep{jp15} and \citep{agp16}, although neither accounts for intraday products with different times to maturity; and while \cite{agp16} model the intraday market, they do not consider storage. \cite{bertrand2019reinforcement, bp19} train a threshold policy for a storage unit on German limit order book data using reinforcement learning. The authors report significant gains for their trading strategy over a greedy strategy. \cite{boukas2021deep} deploy a markov decision process–based modeling framework with a distributed fitted Q-iteration algorithm to optimize to decide when to employ the rolling intrinsic algorithm to trade on the intraday market.

A limited amount of research literature has been devoted to the bidding strategy in the in reserve markets. A small body of work \citep{dimoulkas2016forecasting,kraft2020modeling} has concentrated on the prediction of frequency containment reserve (FCR) prices. The former predicted the SE2 price and volume in the Nordic balancing market using a hidden Markov model. The latter study employed a statistical model and neural networks in a rolling one-step framework to estimate the capacity-weighted average price. The authors of both papers did not propose a bidding strategy based on their forecasts. A bidding strategy for the Nordic FCR is proposed in \citep{div20}, where the trading problem is modeled as a non-linear constrained optimization problem, explicitly taking into account degradation, energy, and balancing costs. \cite{thien17} investigates the profitability of different strategies for the German FCR market and \cite{br16} compares revenues on the Italian primary reserve market with arbitrage strategies on the wholesale market. \cite{astero2020stochastic, 6872593,6021358} consider using electric vehicle batteries in the FCR, while \cite{broneske2017} consider the same problem in the secondary reserve market. 

An important topic when participating in the primary reserve market is the issue of state-of-charge (SoC) recovery after a frequency excursion event. In \citep{zhang16}, a theoretical stochastic dynamic programming framework was proposed to plan and control a BESS that participates in frequency reserves. \cite{xu2014bess} propose a SoC
recovery by trading on the IDM. An online control mechanism was proposed in \citep{div20}.

In the papers reviewed so far, either the balancing market or the IDM are considered. The authors in  \citep{biggins2022trade} propose a bidding strategy in the frequency reserves and day-ahead markets in Great Britain based on machine learning. Their results show that the firm frequency response (FFR) is a stronger source of revenue than the day-ahead market for battery storage and highlight that the simultaneous performance in the FFR and the day-ahead market increases profits. Using machine learning-based price prediction and uncertainty metrics, three bidding strategies are devised to trade in different frequency reserves in \citep{kempitiya2020artificial}. Although multiple power markets were considered in the aforementioned papers, they do not consider the IDM in their strategies. 
    
In this paper, we propose the first approach that combines a realistic representation of a continuous intraday market with participation on a frequency control market. We use a version of the rolling intrinsic (RI) strategy to make trading decisions on the IDM and recovery SoC and machine learning classifier to decide how much capacity to commit to which market. The aim of the overall approach is to define an algorithm that can be used for high-frequency automatized decision-making.

In particular, the main contributions of the paper are:
\begin{enumerate}
    \item We propose a mixed-integer linear programming version of the intrinsic policy that can handle orderbook-based continuous trading. Furthermore, we detail a rolling intrinsic algorithm that, given commitments on the primary reserve market, coordinates the repeated re-computation of the intrinsic strategy with the necessary SoC recovery necessary for participation on the primary reserve market.

    \item We devise a machine learning classifier that decides how much of the available battery capacity to commit on the reserve market. The decision is based on features that are available at the time of FCR bidding and the aim is to predict the times when the RI is more profitable than the FCR market and to optimally split the BESS' capacity between the two markets. 

    \item We present a comprehensive and realistic out-of-sample study that benchmarks the proposed strategy against a battery of alternatives using two years of real market data. Using detailed historical data from these two markets, our results show that using our algorithm to combine FCR and IDM trading returns up to $4$\% more revenue than delivering FCR only and delivers $96$\% more than trading in the continuous intraday market only. Our algorithm clearly outperforms naive strategies and is only $4$\% worse than a clairvoyant strategy that always chooses the right mix between the two markets.
\end{enumerate} 

The rest of the paper is structured as follows: In Section \ref{sec:setting}, we will discuss the IDM and FCR markets with a focus on the European and, more specifically, the German situation. We outline market rules and describe interactions in the form of constraints that bids in one market imply for the other. Section \ref{sec:methods} describes our implementation of the RI method, as well as the classifier that decides between the two markets. Section \ref{sec:numerical_results} is devoted to the results of our case study. We first give some insight into stylized facts related to strategies in order to motivate our choices for the out-of-sample study and then present a detailed comparison of our proposed strategy with a set of benchmarks for the years 2023 and 2024. Section \ref{sec:conclusion} concludes the paper.

\section{Setting \& Markets} \label{sec:setting}
\begin{figure}[t]
    \centering
    \includegraphics[width=1.0\textwidth]{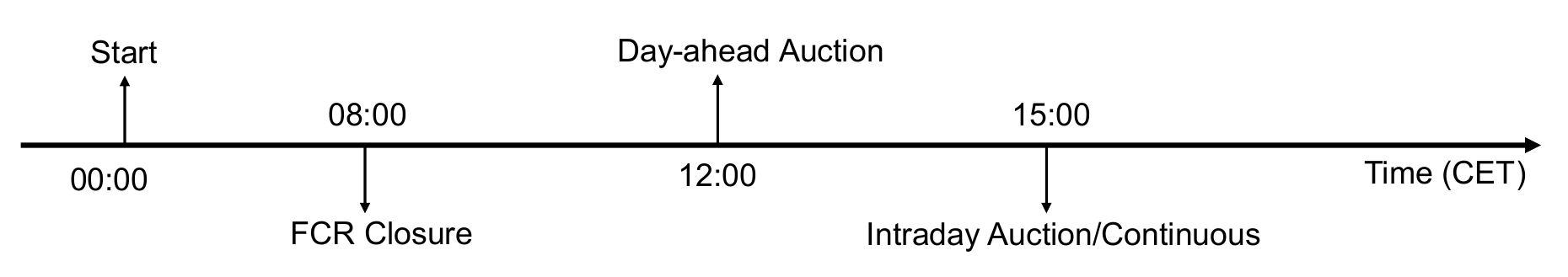}
    \caption{\label{fig:trading_time} Order of events on any given trading day for EPEX day-ahead and intraday markets and the FCR auctions in Germany.}
\end{figure}

In this section, we describe the relevant features of the FCR market and the IDM. To fix ideas and since our numerical study in Section \ref{sec:numerical_results} uses Germany as an example, we focus on the German situation as a typical example of a European market design.

In Figure \ref{fig:trading_time}, we show the order and clearing times for the relevant markets. On any given day, the FCR auctions close at 8 am. The IDM market opens at 3 p.m. for the hourly products and at 4 p.m. for the quarter hourly products. Hence, in our setting, we have to first decide about the bids on the FCR market before the trading on the IDM starts. In the following, we give a detailed description of the relevant aspects for both markets. Table 1 gives a first impression in terms of size, bidding frequency, and traded volumes.

\begin{table}[t] \centering
    \begin{tabular}{lllllllll}
    \toprule
                        & \multicolumn{4}{c}{\textbf{FCR}} & \multicolumn{4}{c}{\textbf{Intraday}} \\
                        & 2020 & 2021 & 2022 & 2023 & 2020 & 2021 & 2022 & 2023\\
                        \cmidrule(lr){2-5} \cmidrule(lr){6-9}
        \textbf{Traded quantity} & 573 & 562 & 555 & 570 & 77 & 85 & 92 & 120\\
        \textbf{Bidding frequency} & \multicolumn{4}{c}{Daily, 4-hour blocks} & \multicolumn{4}{c}{Continuous}\\
        \textbf{Pricing mechanism} & \multicolumn{4}{c}{Uniform price auction} & \multicolumn{4}{c}{Order book based}\\
    \bottomrule
    \end{tabular}
    \caption{\label{tab:market_overview} Daily volumes auctioned in the German FCR market (in MW) from Bundesnetzagentur (see \url{https://data.bundesnetzagentur.de/}), combined EPEX intraday annual trading volumes (in TWh) for the CWE region (AT, BE, DE/LU, FR, NL) taken from the EPEX Spot annual reports \citep{epex_annual_reports}.}
\end{table}

\subsection{The Market for Frequency Containment Reserve} \label{ssec:fcr}
Maintaining a balance between electricity production and consumption is crucial for the stability of the power grid. In a situation where production and consumption do not exactly match, the grid frequency deviates from the nominal frequency, which could damage electronic devices, lead to a failure of components, and may result in selective power cuts or even system-wide blackouts. The FCR is a service organized by the transmission system operators with the goal of recovering from frequency deviations and outages by stabilizing frequencies near the nominal value.

The four German transmission system operators hold a joint procurement auction for FCR capacity each day at 8am for the following day. The day is split into six blocks of four consecutive hours, so-called EFA (\emph{electricity forward agreement}) blocks. There are six simultaneous  uniform-price procurement auctions, one for each EFA block of the following day, in which companies offer their capacity. The minimum bid size in the auction is $1$ MW with increments of $1$ MW \citep{figgener2022influence}. Divisible and indivisible bids are both allowed, and the maximum size of an indivisible bid is $25$ MW. 

\begin{figure}[t]
    \centering
    \includegraphics[width=0.75\textwidth]{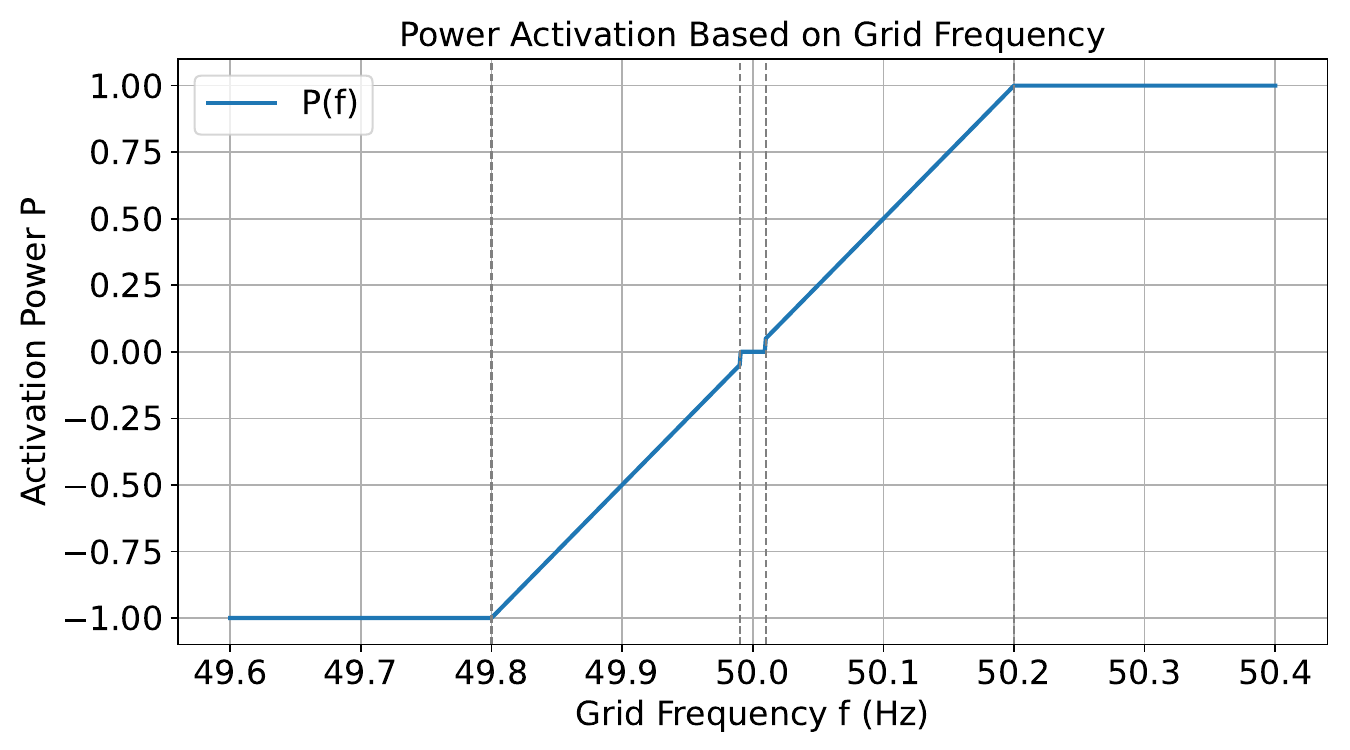}
    \caption{\label{fig:FCR_activation} Required activation of reserve power for a bid of $P_{bid}$ MW as a function of the frequency deviation $\Delta f$.}
\end{figure}

Successful bidders for a specific EFA block are required to instantaneously provide an amount of power proportional to the size of the accepted bid and the deviation of the grid frequency $f$ from the nominal frequency $f^n$ ($50$ Hz in Germany), defined as $\Delta f := f - f^n$. More specifically, FCR market participants have the obligation to provide upward and downward regulations if the absolute deviation $\Delta f$ of the grid frequency exceeds a deadband of 0.01 Hz: If $\Delta f< 0$ upward regulation is provided, that is, additional energy must be fed into the grid. In contrast, if $\Delta f> 0$, that is, the frequency exceeds the nominal frequency, the FCR providers are required to increase energy consumption. The activation $P$ resulting from an accepted bid of $P_{\text{bid}}$ MW is defined as (see Figure \ref{fig:FCR_activation}),
\begin{equation}\label{eq:FCR_activation}
    P= \begin{cases}
    		0, & \text{if $|\Delta f| \leq$ 0.01 Hz }\\
            \frac{\Delta f}{0.2}\cdot P_{\text{bid}}, & \text{0.01 Hz $ < |\Delta f| \leq$ 0.2 Hz}\\
            \frac{\Delta f - 0.2}{|\Delta f -0.2|}\cdot P_{\text{bid}}, & \text{ $|\Delta f| > $ 0.2 Hz},
    	\end{cases}    
\end{equation}
where a positive $P$ implies that energy has to be absorbed from the grid while a negative sign indicates that additional energy has to be provided~\citep{thien17}.

Firms must deliver at least $50$\% of activation within $15$ seconds after receiving the signal $\Delta f$ and should be fully ramped up/down after $30$ seconds. Furthermore, firms have to guarantee an energy capacity large enough to cover the full activation $P_{\text{bid}}$ for at least $15$ minutes per incident \citep{figgener2022influence}. Failure to offer the promised balancing service results in a penalty. If an imbalance persists for more than $30$ seconds, the TSO dispatches secondary reserves that relieve the providers of primary reserve. If there is still a positive $|\Delta f|$ after $5$ minutes, the manually controlled tertiary reserve and possible redispatch mechanisms are used to return the system to equilibrium.

In order to guarantee the 15-minute criterion, the energy stored in the BESS has to remain in the range of $[0.25 P_{\text{bid}}, \bar{E}-0.25 P_{\text{bid}}]$, where $\bar{E}$ is the battery's energy capacity. However, there may be more than one frequency excursion event in an EFA block and participants are obliged to provide the service for all incidents. Therefore, after each FCR activation, the providers must restore their SoC to an acceptable range that allows them to service the next event. The above also implies an upper bound $0.25P_{\text{bid}} \leq 0.5\bar{E}$ for the bid power, which is, however, rarely of practical relevance since the power capacity (in MW) usually does not exceed the energy capacity (in MWh) for BESS. 

Clearly, at the time of bidding capacities for the FCR market, the direction and magnitude of $P$ is uncertain. For storage, this has two implications: Firstly, the operators have to ensure that the SoC is in a range where it is guaranteed that they can honor their commitments made on the FCR. Secondly, although cumulative activations are, in most periods, close to zero, it often happens that the SoC at the end of a 4-hour EFA block significantly differs from the initial SoC, i.e., that activations do not average out. This effect is further exacerbated by the fact that battery efficiency losses (charging and discharging inefficiencies) result in a net energy loss even if positive and negative FCR activations within an EFA block cancel out on average.

\subsection{The Intraday Market} \label{ssec:idc}
The European intraday market framework is designed around the principle of market integration and convergence toward a single European electricity market. The European Commission established a target model to integrate all intraday markets based on continuous trading through the Commission Regulation (EU) 2015/1222 establishing guidelines on \emph{capacity allocation and congestion management}. This integration is facilitated by the Cross-Border Intraday (XBID) platform, which operates with a shared order book that enables market participants to trade electricity continuously across European borders, provided that sufficient cross-border transmission capacity is available. The aim of this \emph{single intraday coupling} is to create a single EU cross-zonal intraday electricity market where buyers and sellers can work together across Europe to trade electricity continuously on the day the energy is needed \citep{EC2015CACM}. It serves as a complementary market to the day-ahead market and is used to continuously buy or sell electricity to manage short-term deviations from sudden changes in energy supply and demand. 

The intraday market in Germany is mainly organized by the European Power Exchange (EPEX), the largest intraday power exchange in Europe. In recent years, the intraday market has been growing steadily, as can be seen from the traded volumes in Table \ref{tab:market_overview}. The market is organized as an order book-based continuous trading market that features hourly, half-hourly, quarter-hourly, and block products. Currently, market participants in Germany can trade until $5$ minutes before physical delivery in their control areas and until up to $30$ minutes before delivery in a combined national market \citep{kuppelwieser2023intraday}. 

Each buy and sell order on the intraday market for a given product contains basic information about quantity, limit price, and validity time. A \emph{market order} is cleared immediately against the best available order in the limit order book (LOB), while a limit order is only executed with matching orders on the other side of the market up to a certain price (the limit). If this is not possible, the order is kept in the LOB until its \emph{end validity date} to be cleared with future orders. If the quantities of two matched orders do not agree, the order with the higher order quantity is only partially cleared and remains in the order book with the remaining quantity. The minimum bid volume is $0.1$ MW, and the price per MWh ranges from $-9999$ to $9999$ euros.

Market participants can add the usual order qualifiers such as \emph{immediate-or-cancel} (IOC) or \emph{fill-or-kill} (FOK). Additionally, \emph{iceberg} orders are allowed for which only a fraction of the order quantity is visible to other market participants. As soon as the visible quantity is cleared, the next part of the order is automatically placed in the limit order book.

\begin{figure}[t]
    \centering
    \includegraphics[width=0.3\textwidth]{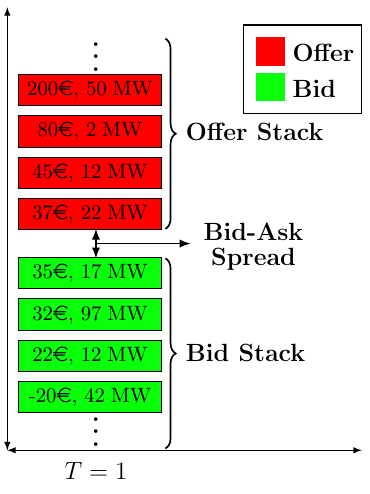} \hspace{.08\textwidth}
    \includegraphics[width=0.6\textwidth]{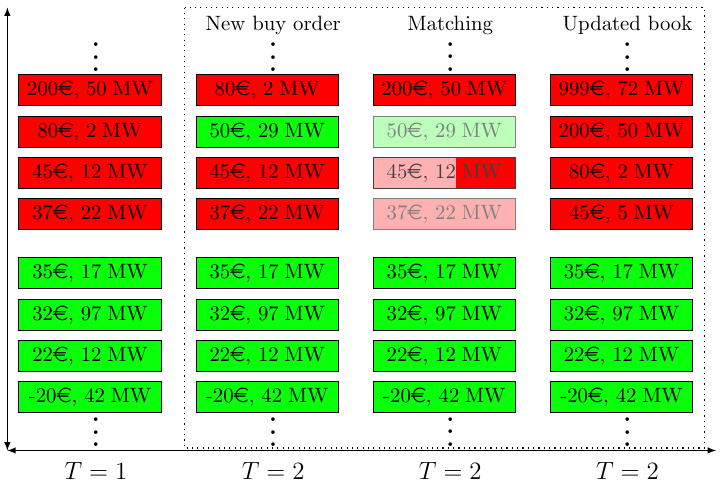}
    \caption{\label{fig:cont_trading} Illustration of the limit order book clearing mechanism. Source \citep{GrKuWo24}.}
\end{figure}

The basic mechanism of continuous trading is illustrated in Figure \ref{fig:cont_trading} by a concrete example \citep[taken from][]{GrKuWo24}: In the left panel, the state of the order book at $T=1$ is displayed with the orders sorted according to their limit price and organized into a  \emph{bid stack} and \emph{ask stack}. The state of the LOB changes with the placement of a new order, with the modification of an order, and at the end-validity-time of an active order. The limit price of the order with the lowest sell price is called \emph{best-ask}, while the order with the highest buy price defines the \emph{best-bid}, and the difference between the two prices is the \emph{bid-ask-spread}. In the right panel, the clearing of a new hypothetical buy order is illustrated: A new buy order with a price of 50\euro \, is added to the book at $T=2$ and is then cleared against the cheapest possible offers until either the whole order is fulfilled (as is the case in the figure) or there are no offers with lower prices left. In this example, 22\;MW out of 29\;MW are cleared against the sell order with price 37\euro \, and the remaining $29\text{\;MW} - 22\text{\;MW} = 7\text{\;MW}$ are cleared against the sell order with price 45\euro. The remaining quantity of $12\text{\;MW} - 7\text{\;MW} = 5\text{\;MW}$ of the latter order stays in the order book. Note that the clearing is instantaneous, i.e., columns $2$ and $3$ in the right panel are purely illustrative and do not correspond to market states that can be observed by traders.

\section{A Joint Optimization Method for Intraday FCR Trading} \label{sec:methods}
Participating in both the FCR and the IDM involves (i) bidding in the FCR auction one day ahead of delivery and (ii) trading in the IDM while honoring the battery's FCR commitments. Finding an optimal combined bidding strategy for the FCR and the IDM thus involves solving a fairly complicated stochastic optimization with a large number of intraday trading stages that model the possibility of frequent rebalancing. Solving this problem exactly is computationally intractable.

We therefore propose a heuristic policy that selects one of finitely many FCR bidding strategies and then uses the RI policy to solve the intraday bidding problem. To that end, we frame the problem of selecting an optimal FCR strategy as a classification problem that picks the strategy with the highest overall expected profits.

We discuss the strategy space for the FCR market in Section \ref{ssec:FCR_strategies}, a version of the rolling intrinsic policy adapted to continuous trading in Section \ref{ssec:RI}, and finally the classification problem for the different strategies in Section \ref{ssec:classification_task}.

\subsection{Overview of Joint Bidding Model} \label{ssec:FCR_strategies}
For each EFA block $i$, suppliers can bid the price (EUR/MW) as the requested compensation for the service. In this paper, we assume that the bidding strategy only includes the quantity and that the bidding price is zero. Due to the pay-as-cleared mechanism, even with the bidding price being zero, providers still get the uniform market clearing price as their final remuneration. 

More specifically, we use a vector $X = (X_1, X_2, X_3, X_4, X_5, X_6)$ with six components to represent a bid strategy in the six FCR markets, where $X_i$ denotes the capacity (in MW) that is committed to the market $i$. For a battery with capacity $\bar X$, this implies that in principle $(\bar X - X_i)$ MW are available for trading on the intraday market in the time period covered the EFA block $i$. Hence, the FCR strategy determines the feasible bids on the intraday market. 

The regulation of the FCR market in Germany stipulates that a storage can provide the maximum up-regulation or down-regulation of the full reserve bid $X_i$ for at least $15$ minutes without interruptions (see Section \ref{sec:setting}). Therefore, the storage level must be in $[\nicefrac{X_i}{4}, \bar E - \nicefrac{X_i}{4}]$ at any time during the period covered by the market $i$. Furthermore, we limit the maximum FCR bid to 80\% of the asset's power capacity. Note that this restriction is not required by market rules but is based on backtesting evidence, which shows that larger bids often caused SoC management issues. This assumption is reasonable for BESS with a power-to-energy ratio, as considered in this study (approximately 1).

As described in Section \ref{ssec:fcr}, admissible bids are in whole MW with a minimum bid size of $1$ MW. Hence, for a storage with capacity $\overline X$, there are $\lfloor 0.8 \cdot \overline{X} + 1\rfloor ^6$ different admissible bidding strategies for the FCR markets of a single day in our setting.  

If $X_i = 0$, the lower limit on stored energy in the corresponding time periods $t$ covered by market $i$ equals $\underline{\alpha} \times \bar E$ while the upper limit is $\overline{\alpha} \times \bar E$. These limits are in place to preserve battery health and to avoid complete discharge as well as overcharging. In case that $X_i > 0$ these bounds tighten according to the rules outlined above. Hence, for any time $t$ in the time span covered by market $i$, we get that the state of charge $c_t$ has to be in the interval
\begin{equation} \label{eq:energy_bounds}
    [\underline{c}_t, \overline{c}_t] := \left[\max(\underline{\alpha} \times \bar E, \nicefrac{X_i}{4}), \min(\overline{\alpha} \times \bar E, \; \bar X - \nicefrac{X_i}{4})\right].
\end{equation}
This limitation is incorporated into the RI as a constraint as described in the next section. If the FCR activation causes the battery charge to exceed the boundary, the constraint requires the RI to immediately restore the SoC to acceptable limits by corresponding IDM trades.

Although positive and negative deviations from activations on the FCR market tend to cancel out, participation in the market can produce a slight drift in the state of charge of a storage. This is acerbated by the efficiency losses of storage. The energy drift in the time interval $[t, t+\Delta t]$ is dependent on fluctuations of the national grid frequency and can be expressed as
\begin{equation} \label{eq:energy_drift_estimator}
    D_{t,t+\Delta t} = \int_t^{t+\Delta t} P(\Delta f_s) \eta(P(\Delta f_s)) \; ds \approx \frac{\Delta t}{K} \sum_{k=1}^{K} P(\Delta f_{t+k\frac{\Delta t}{K}}) \eta(P(\Delta f_{t+k\frac{\Delta t}{K}})),
\end{equation}
where $P$ is defined in \eqref{eq:FCR_activation} and $\Delta f_s$ is the deviation of the national frequency at time $s$. Furthermore, $\eta$ maps the positive and negative deviations to the charging and discharging efficiency, respectively, and $K$ is a positive integer defining the discretization grid in the approximation of the integral by a Riemann sum.

When making a bidding decision on the FCR market, the BESS trades off revenues for every EFA block with the revenues that can be potentially earned on the intraday market. The total daily profits $\pi$ are the sum of the profits $\pi_{FCR}$ from bidding on the FCR market and the profits $\pi_{IDM}$ from trading on the intraday market.
\begin{equation} \label{eq:daily_profits}
    \pi(X) = \underbrace{\sum_{k=1}^6 P_{FCR}^{k}X_k\vphantom{\sum_{j\in O_b}}}_{\pi_{FCR}(X)} 
    + \pi_{IDM}(X),
\end{equation}
where $P_{FCR}^{k}$ is the clearing price of the FCR for the EFA block $k$ and the profits profits on the IDM are a function of $X$ as will be discussed in the next section. Since neither $P_{FCR}^k$ nor $\pi_{IDM}(X)$ is known at the time of bidding on the FCR market, $\pi(X)$ is a random quantity.

Therefore, in order to select an FCR strategy for the six markets of one day, we aim to solve the multi-stage stochastic optimization problem 
\begin{equation} \label{eq:max_exp_profits}
    \max_{X \in \Xc} \; \Ed[\pi(X)],
\end{equation}
where bids $X$ are the first-stage decisions, $\Xc$ is the feasible set, and the bids on the IDM represent the recourse decisions. Since the intraday market is continuous, the problem has an excessive amount of stages and therefore solving \eqref{eq:max_exp_profits} exactly is computationally intractable.

\subsection{The Continuous Rolling Intrinsic Strategy for Intraday Trading} \label{ssec:RI}
The intrinsic value of a storage asset is the profit that can be realized by exploiting the currently observable price spreads in the market. A static intrinsic strategy therefore determines an optimal set of buy and sell positions based on the forward curve at a single point in time. This strategy focuses on locking in guaranteed profits without speculating on future price movements.

The RI policy, originally introduced in \citep{gray2004towards} for the evaluation of gas storage, is a dynamic extension of the intrinsic value. Starting from an initial SoC and a portfolio of forward positions acquired in previous periods on the intraday market, the RI repeatedly checks for chances of profitable rebalancing by re-solving the intrinsic problem. Although the resulting decisions are still myopic, since the policy at no point trades on anticipated future price changes, the RI clearly represents an improvement over the static intrinsic that does not adapt positions at all. Furthermore, the myopic nature of decisions has the advantage that the RI only enters immediately profitable positions and therefore does not run the risk of accumulating losses based on wrong assumptions about the future. For these reasons and because of its conceptual simplicity and low computational cost, the RI has attained widespread industry adoption.

The classic version of the RI is based on a price taking intrinsic strategy. To be able to employ the algorithm to the continuous intraday market, we adapt the intrinsic algorithm to explicitly take into account the LOB and corrections based on energy drift from FCR market activations when trading. In particular, for every rebalancing decision, the algorithm receives a snapshot of the current state of the order book and subsequently checks whether the positions of forward contracts built up in previous steps can be profitably updated. 

In order to discuss the RI, we first formally define the intrinsic problem for continuous intraday markets, which is solved repeatedly in the process. To that end, we define the set of tradeable contracts $\Tc$ at the time the intrinsic policy is executed, as well as the order book information $\Oc = \bigcup_{t \in \Tc} \Oc_t$, which contains order numbers of all bids and asks currently in the orderbook $\Oc_t$ for contract $t$. In addition, we define the set of contracts that go into delivery on the current day as $\Tc_0$. For an order $i \in \Oc$, we denote by $P_i$ its limit price (in EUR/MWh), by $\sigma_i$ its direction ($-1$ for bid and $1$ for ask), and finally by $Q_i$ the quantity. In order to capture the history of trading decisions, we denote by $c_0$ and $b_t^0$ the energy and power commitments resulting from previous trades where $c_0$ is adjusted for the energy drift that results from call offs on the FCR market and by products that go into delivery (see below). To model the cost of battery degradation, we use a simple linear approximation by multiplying the absolute magnitude of planned charging/discharding by the degradation cost $\kappa>0$ (in EUR/MWh).

Note that the last tradeable contract $t$ is always the last period traded in a day. We fix a terminal storage level $C_T$ at the end of a trading day, i.e., at time $T = \max \; \Tc$. With these preparations in place, we can formulate the intrinsic problem, which decides for every order $i \in \Oc$ about the quantity $q_i$ (in MW) that is matched by the intrinsic policy as follows.
\begin{subequations} \label{eq:RI}
	\begin{align}
		\underset{q_{i},c_t,b_t, b_t^\pm}{\max} \quad & \Delta \sum_{t \in \Tc}\sum_{i \in O_t} P_{i} \sigma_{i} q_{i} - \kappa \Delta \sum_{t \in \Tc} (b_{t}^+ + b_t^-) \label{eq:obj} \\
		\text{s.t.} \quad & 0 \leq q_{i} \leq Q_{i}, & \forall i \in \Oc,\\
		& b_{t}=b_{t}^0 - \sum_{i \in \Oc_t} \sigma_{i}q_{i}, & \forall  t \in \Tc, \label{eq:power} \\
        & b_t = b_t^+ - b_t^-, & \forall  t \in \Tc, \label{eq:b_split}\\
        & b_t^+ \leq \delta_t \overline b_t, & \forall  t \in \Tc, \\
        & b_t^- \leq (1-\delta_t) (-\underline b_t), & \forall  t \in \Tc, \label{eq:b_neg} \\
		& c_{t} = c_0 + \Delta \sum_{a \leq t} \eta_{ch} b_a^+ - \Delta \sum_{a \leq t} \frac{1}{\eta_{dis}} b_a^-, & \quad \quad \forall  t \in \Tc , \label{eq:energy} \\
		& \underline b_t \leq b_{t} \leq \overline b_t, & \forall  t \in \Tc, \label{eq:power_limits}\\
		& \underline c_t \leq c_{t} \leq \overline c_{t} & \forall  t \in \Tc,\label{eq:energy_limits}\\
        & b_t^+, b_t^- \geq 0, & \forall  t \in \Tc\\
        & \delta_t \in \{0,1\}, & \forall  t \in \Tc\\
		& c_T = C_T, \label{eq:end_level}\\
		& \frac{\Delta\sum_{t\in \Tc} (b_{t}^+ + b_t^-)  }{2 \bar{E} } \leq \bar C^0, \label{eq:cycling_limit}
	\end{align}
\end{subequations}
where $\Delta$ is the duration of the delivery period of the traded contracts (e.g., one hour or $15$ minutes), $\overline b_t$, $\overline c_t \in \Rd^+$, $\underline b_t \in \Rd^-$, $\underline c_t\in \Rd^+$ are upper and lower bounds on stored energy and charged discharged power in period $t \in \Tc$ (see discussion above), respectively and $\eta_{ch}, \eta_{dis} < 1$ is the charging/discharging efficiency factor (depending on the direction of the trade). Note that the bounds on power and energy depend on the commitments on the FCR market for the corresponding EFA block (see last section).

The variables $b_{t}$ defined in \eqref{eq:power} model the accumulated power (in MW) of all orders with the delivery period $t$ and depend on the corresponding values from previous trades $b_t^0$ and the decisions in the current problem, where $b_t^+, b_t^-$ are the positive and negative parts of $b_t$, respectively, as modeled by \eqref{eq:b_split} to \eqref{eq:b_neg}, where the constraints make sure that for every product energy is either bought or sold, which is important for contracts with negative prices where simultaneous charging and discharging is potentially profitable, depending on efficiencies and the observed bid/ask spread. Similarly, the variables $c_t$ in \eqref{eq:energy} define the total energy stored in the storage until the end of period $t$. Note that the factor $\Delta$ translates power (in MW) into energy (in MWh). The constraints \eqref{eq:power_limits} and \eqref{eq:energy_limits} represent the limits for power and energy imposed by the physical limitations of the storage as well as commitments on the FCR market. The last constraint \eqref{eq:end_level} requires that the storage level is equal to $C_T$ at the end of the planning horizon. The constraint \eqref{eq:cycling_limit} enforces a daily cycling limit, where $\bar C^0$ is the number of permissible cycles left on the current day, which is updated in the RI Algorithm \ref{alg:RI}. Finally, the first term of the objective function \eqref{eq:obj} models the profits from trading, while the second term yields the planned losses from battery degradation. 

Note that while on the intraday market different types of contracts are traded (hourly, half-hourly, quarter-hourly), we only consider one type in the above formulation. Clearly, at the expense of a more involved notation, the problem could easily be extended to cover different types of contracts. However, for the sake of simplicity and because we only use one type of contract in our numerical study, we refrain from this complication.

The intrinsic problem \eqref{eq:RI} is the main building block of the rolling intrinsic policy, which is detailed in Algorithm \ref{alg:RI}. The RI reoptimizes a given position either periodically or every time there is a change in the LOB. The policy is myopic in the sense that trades are only executed if the resulting immediate profit is positive. This leads to profits that are guaranteed to be non-negative and a strategy that is not able to anticipate future prices foregoing potential profits of more speculative strategies but also eliminating any downside risks.
\begin{algorithm}[t]
\small                   
\SetNoFillComment
\KwData{Storage level $C_0$, planned final storage level $C_T$, $\Tc$, minimal bid size $\delta$}

$\forall t \in \Tc: b_{t}^0 \gets 0$, $c_0 \gets C_0$\;
$\pi \gets 0$, $\Tc^- \gets \Tc$\;

\While{$\Tc \neq \emptyset$}{
  \tcc{Correct for drift and calculate realized battery degradation cost}
  Set $\Delta\Tc \gets \Tc^- \setminus \Tc$, fetch order books $\Oc_t$\ and FCR SoC drift $D$ since last optimization\;
  $c_0 \gets c_0 – D + \Delta \sum_{t \in \Delta \Tc} \eta_{ch} b_t^+ - \Delta \sum_{t \in \Delta \Tc} \frac{1}{\eta_{dis}} b_t^-$\;
  $\pi \gets \pi - \Delta\kappa \sum_{t \in \Delta\Tc} |b_{t}^0|$\;

  \tcc{Update cycling limit}
  \uIf{a new day has started since the last solve}{
    $\bar C^0 \gets N^{cycles}$ \;
  }
  \Else{
     $\bar C^0 \gets \bar C^0 - \frac{\Delta \sum_{t \in \Delta \Tc} |b_t|}{2\bar{E}}$ \;
  }
  \tcc{Initialize new contracts}
  $b_t^0 \gets 0$ for all $\Tc\setminus \Tc^-$ \; 
  \tcc{Solve the intrinsic and round results to implementable trades}
  Solve the intrinsic problem resulting in optimal $(b_t)_{t\in\Tc}$, $(c_t)_{t\in\Tc}$ and $(q_i)_{i\in\Oc}$\;

  \If{intrinsic is feasible}{%
    \relax $q_i \gets \operatorname{round}(q_i/\delta)\times\delta, \quad \forall i\in\Oc$\;
    $\pi \gets \pi + \Delta \sum_{i\in\Oc} P_i\sigma_i q_i$\;
    $b_t^0 \gets b_{t}^0 - \sum_{i\in\Oc_t} \sigma_i q_i,\quad \forall t\in\Tc$\;
  }

  \tcc{Prepare next iteration}
  $\Tc^- \gets \Tc$, wait for next trading time, and update $\Tc$\;
}
$\pi \gets \pi - \Delta\kappa \sum_{t \in \Tc} |b_t|$\;

\vspace{0.3cm}
\caption{\label{alg:RI}Rolling‐intrinsic algorithm for one day of trading on the intraday market.}
\end{algorithm}

Note that the RI profits are not taken from the objective of the problem \eqref{eq:RI}, but calculated outside of the optimization problem in Algorithm \ref{alg:RI}. The reasons for this are twofold: Firstly, to not count battery degradation costs for the same periods multiple times, the degradation cost is accounted for in lines 6 and 20 for the periods that are no longer traded and go into physical delivery. Secondly, some of the trades found by the linear program \eqref{eq:RI} may not be feasible, since they are not forced to be multiples of the minimum bid size $\delta$. Note that this happens only for the most expensive accepted ask and the cheapest accepted bid per period. We correct for these inaccuracies in line 14.

On a related note, we remark that the initial schedule $(c_t^0)_{t\in \Tc}$ for any of the intrinsic solves need not be physically feasible due to the rounding after the solution of the last problem, as well as due to the energy drift that occurs between two consecutive runs of the intrinsic policy. Everytime \eqref{eq:RI} is solved, these violations are corrected. If there are not enough orders in the LOB for correction, the problem \eqref{eq:RI} is potentially infeasible. Hence, the if-statement in line 13. If such as situation persists over a longer period of time, we theoretically could end up with a physically infeasible schedule that has to be corrected on the balancing market. However, in Section \ref{sec:numerical_results}, we demonstrate that this case does not occur in practice for sufficiently conservative limits to storage operation.

\subsection{Strategy selection as a classification task} \label{ssec:classification_task}
In this section, we describe how we approximate \eqref{eq:max_exp_profits}. The main idea is to simplify the problem by first deciding among the finitely many feasible $X \in \Xc$ and then relegating the recourse decisions to the rolling intrinsic alogrithm.

To be able to choose among the strategies in $\Xc$, we fit a classification model with the objective of picking the strategy $X \in \Xc$ that, paired with the RI trading strategy in the IDM, produces the highest average profits. We base the predictor on a supervised learning method that uses a set of features and is calibrated using historical data. More specifically, for every day $d \in \Dc$ in our training data, we have a feature vector $f_d$ and profits $\pi_d(X)$ as calculated in \eqref{eq:daily_profits} dependent on the FCR strategy and on IDM trading by the RI.

Thus, we frame the optimal strategy selection for the FCR market as a classification task where the features $f_d$ are used to forecast the labels $X_d^*$ with $X_d^*$ the strategy $X$ that maximizes trading profits on day $d$, i.e.,
$$ X_d^* = \argmax \set{\Ed[ \pi_d (X)]: X \in \Xc }.$$
To do so, we fix a hypothesis class $\Hc$ such that $h: \Fc \to \Xc$ for every $h \in \Hc$, where $\Fc$ is the feature space and solve 
\begin{equation} \label{eq:classification_problem}
    h^* = \argmax_{h\in \Hc} \frac{1}{|\Dc|} \sum_{d \in \Dc} L(X^*_d, h(f_d)),
\end{equation}
with $L$ the cross entropy loss.

Note that in order to calculate the labels of the above problem, we have to evaluate the profits $\pi_d(X)$ for every strategy $X \in \Xc$ for all days $d \in \Dc$ in the training data. However, evaluating IDM trading with order book data for every of the $\lfloor 0.8 \cdot \overline{X} + 1\rfloor^6$ candidate FCR strategies is computationally expensive. For this reason and in order to limit the potential class imbalance in \eqref{eq:classification_problem}, we restrict the set $\Xc$ of possible strategies. 

In order to do this, we proceed in two steps. In a first step, we limit the set $\Xc$ to a reduced set of strategies $\Mc \subseteq \Xc$ based on our analysis of the interaction between the revenues that can be generated in the two markets in Section \ref{sec:numerical_results}. We evaluate profits for all days $\Dc$ and all strategies $X \in \Mc$. To further reduce class imbalance, we reduce the strategy space to $\Sc \subseteq \Mc$ base strategies. An approach to do this would be to simply select a fixed set of strategies from $\Mc$ that yield the highest average profits. However, this does not necessarily yield an optimal subset of strategies, since rather similar strategies might be chosen instead of designing $\Sc$ to consist of complementary strategies.\footnote{To see this consider the case where $|\Mc| = 3$ and $|\Sc|=2$ and where there are two strategies $X^1$ and $X^2$ that perform best on average but are rather similar as well as a third strategy $X^3$ which yields good results on those days where $X^1$ and $X^2$ have poor performance. In this case, it is clearly advantageous to choose $\Sc = \{X^1, X^3\}$ as a set of complementary strategies which contains a good strategy for every day rather than simply choosing the best two strategies.} 

In order to find a well performing  yet small strategy pool set of size $S < |\Mc|$, we pick a subset $\Sc \subseteq \Mc$ producing the highest overall profit on the training data under the hypothesis of perfect classification, i.e., under the assumption that for every day the best strategy $X \in \Sc$ is chosen by the classifier. We do this by solving the following mixed integer linear program
\begin{subequations}
\label{eq:best_group_selector}
    \begin{align}
        \underset{z_X,w_{Xd}}{\max} \quad & \sum_{X \in \Mc} \sum_{d \in \Dc} w_{Xd}\pi_{d}(X) \\
        \text{s.t.} \quad & z_X \in \{0,1\}, & \quad X \in \Mc \\
        & w_{Xd} \in [0,1], & \quad \forall X \in \Mc, \; \forall d \in \Dc\\
        & w_{Xd} \leq z_{X}, & \quad \forall X \in \Mc, \; \forall d \in \Dc \label{eq:relationship_w_z}\\
        & \sum_{X\in \Mc} w_{Xd} =1, & \quad \forall d \in \Dc \label{eq:only_one_strategy}\\
        & \sum_{X \in \Mc} z_X=S. \label{eq:only_S_strategies}
    \end{align}
\end{subequations}
In the above problem, the variables $z_X$ determine whether a strategy is in $\Sc$ and constraint \eqref{eq:only_S_strategies} enforces that only $S$ strategies can be chosen. The variables $w_{Xd}$ model whether a strategy $X$ is chosen for day $d$. Note that since the objective is linear in $w_{Xd}$ the optimal values of $w_{Xd}$ will be in $\{0,1\}$ and the sum in \eqref{eq:only_one_strategy} ensures that only one strategy is chosen per day. Clearly, if $z_X = 0$, $w_{Xd} = 0$ for any day $d$ due to \eqref{eq:relationship_w_z}. Given the optimal solution of the above problem, we define
$$ \Sc = \set{X \in \Mc: z_X = 1}$$
and define our labels as $X_d \in \argmax \set{ \pi_d(X): X \in \Sc }$.

In order to solve the ensuing classification problem based on the data $(X_d, f_d)_{d \in \Dc}$ as our training set, we use \emph{XGBoost} \citep{chen2016xgboost}, a gradient boosting method based on decision trees. Ensemble methods such as gradient boosting are the current state of the art for tabular data and often outperform other methods \citep[e.g.][]{shwartz2022tabular, tabular_data} and specifically XGBoost has shown remarkable performance in machine learning competitions since its inception and offers high computational efficiency, the ability to manage class imbalance, and has powerful feature selection capabilities. 

\section{A Numerical Case Study} \label{sec:numerical_results}
This section discusses a numerical out-of-sample study that puts the method developed in the previous sections to the test. Throughout we assume a 10MW/10MWh battery, i.e., a battery with a power capacity of 10MW and an energy capacity of 10MWh and use $\underline{\alpha} = 0.01$ and $\overline{\alpha} = 0.985$ to define SoC limits. We use historical limit order books from the German continuous intraday market from EPEX Spot and historical FCR clearing prices from Regelleistung.net from 01/01/2023 until 30/09/2024 for our study.

We begin by investigating some static FCR strategies $X$, which means a constant allocation of battery power to the FCR and IDM markets for every day of the observation period. We show that, contrary to intuition, revenue per market does not change linearly with the allocated battery power. We also explain what drives the observed relationship. Having built some intuition around the optimal allocation of battery power between FCR and IDM, we then move to the results of the dynamic FCR strategy as a classification problem.

\subsection{Decreasing Intraday Returns}
We start by discussing the fact that per MW revenues in the IDM are a decreasing function of the amount of battery capacity left for intraday trading. There are essentially two reasons for this: Firstly, limited liquidity of the IDM favors strategies with less capacity, since these do not experience a large price response when trading \citep{kuppelwieser2021liquidity}. Secondly, the less capacity is committed on the FCR market the higher the duration of the remaining storage that can be traded on the intraday market, which tends to decrease profits per unit of capacity. We will discuss this less obvious aspect below.

The battery power allocated to the FCR market determines not only the battery power left on for IDM trading, but also the energy capacity of the battery still available for use on the IDM market. In particular, the FCR regulation states that the battery must at all times be able to delivery the committed power for at least $15$ minutes in both directions (see discussion in Sections \ref{sec:setting} and \ref{sec:methods}). In our case this means that with a commitment of $8$ MW in the FCR market the 10MW/10MWh battery has to maintain a charge level between $2$MWh and $8$MWh. This leaves $2$MW and $6$MWh of flexibility on the IDM market, increasing the energy to power ratio from $1$ (of the original 10MW/10MWh battery) to $3$ on the IDM market.

\begin{figure}[t]
    \centering
    \includegraphics[width = \textwidth]{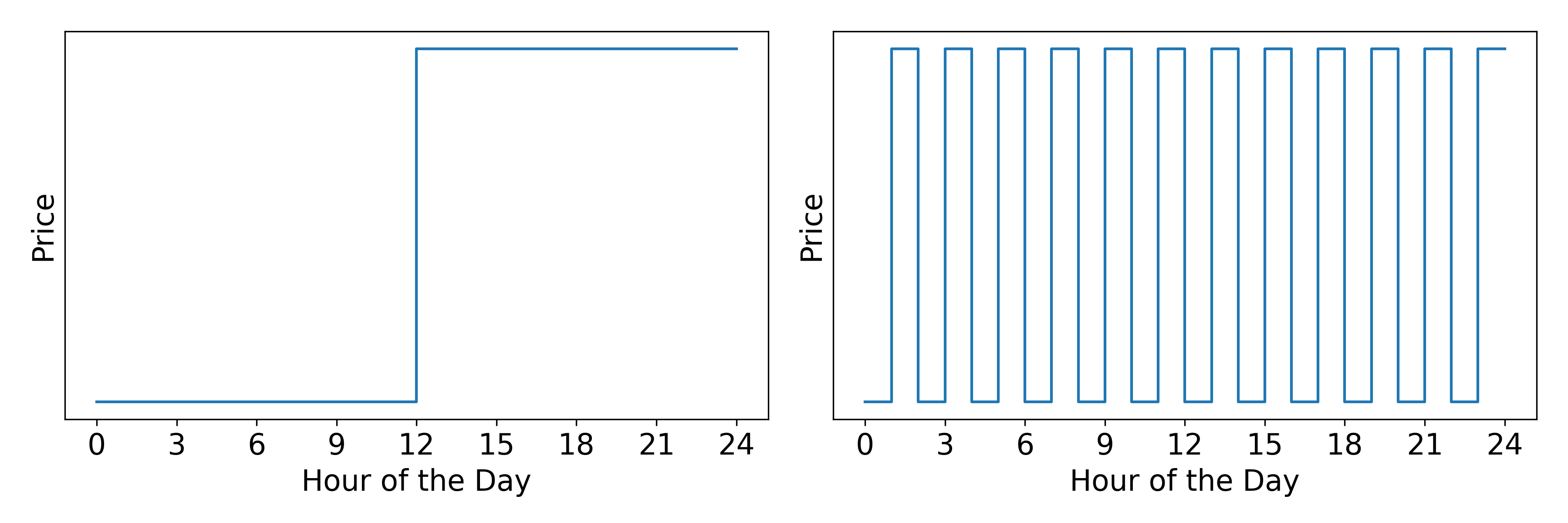}
    \caption{\label{fig:price_patterns} Two extreme price patterns for one day.}
\end{figure}

The increase in total profits from intraday trading when duration changes depends on the price patterns. The two most extreme cases are depicted in Figure \ref{fig:price_patterns}: The left panel shows a situation where the prices in the first $12$ hours of the day are low, enabling the storage to continuously charge in that time and sell back the energy in the second half of the day for high prices. In this case, a perfectly efficient battery with a duration of $12$ hours is optimal. Contrast this with the situation in the right panel, where prices alternate between high and low prices every hour and the storage would only charge for $1$ hour before discharging again and therefore durations above $1$ do not yield additional profits.

\begin{figure}[t]
    \centering
    \includegraphics[width=0.45\textwidth]{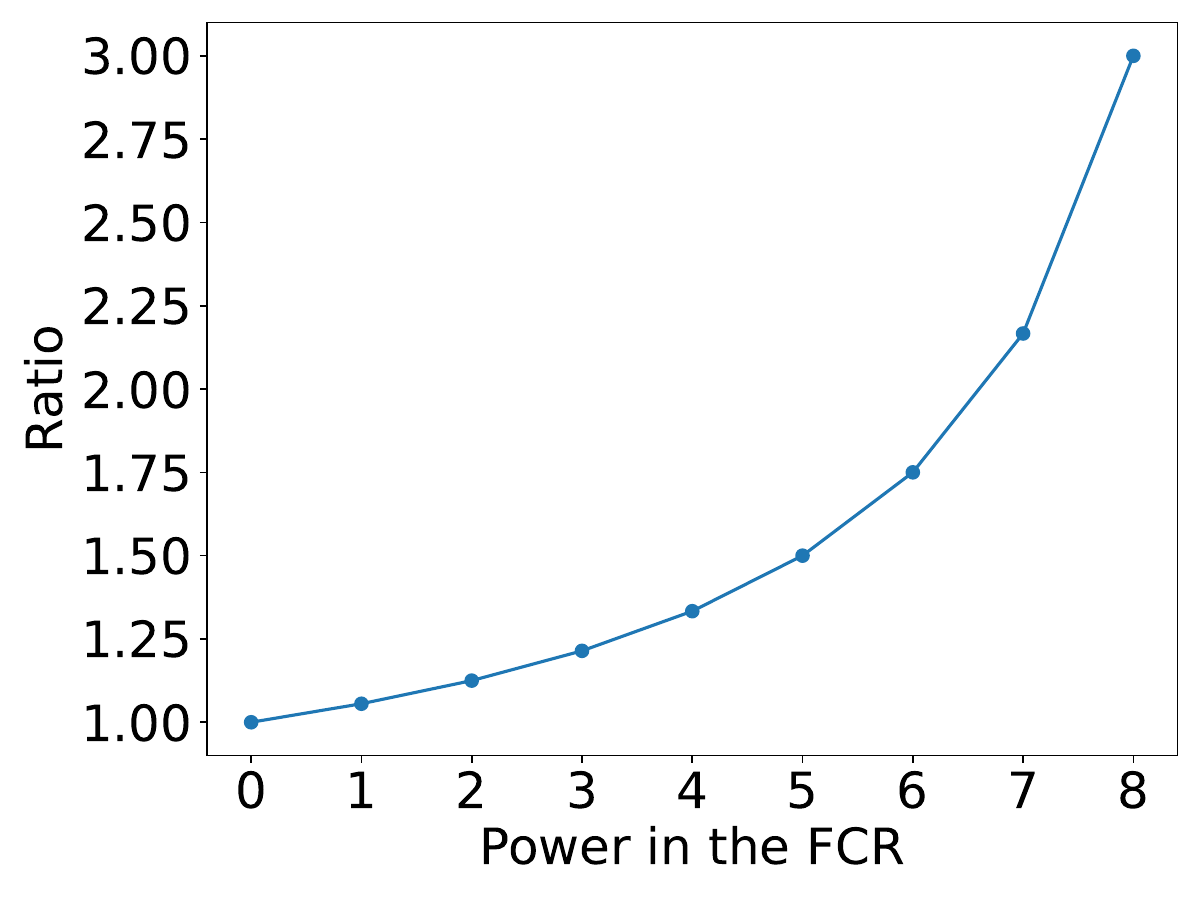}
    \includegraphics[width=0.45\textwidth]{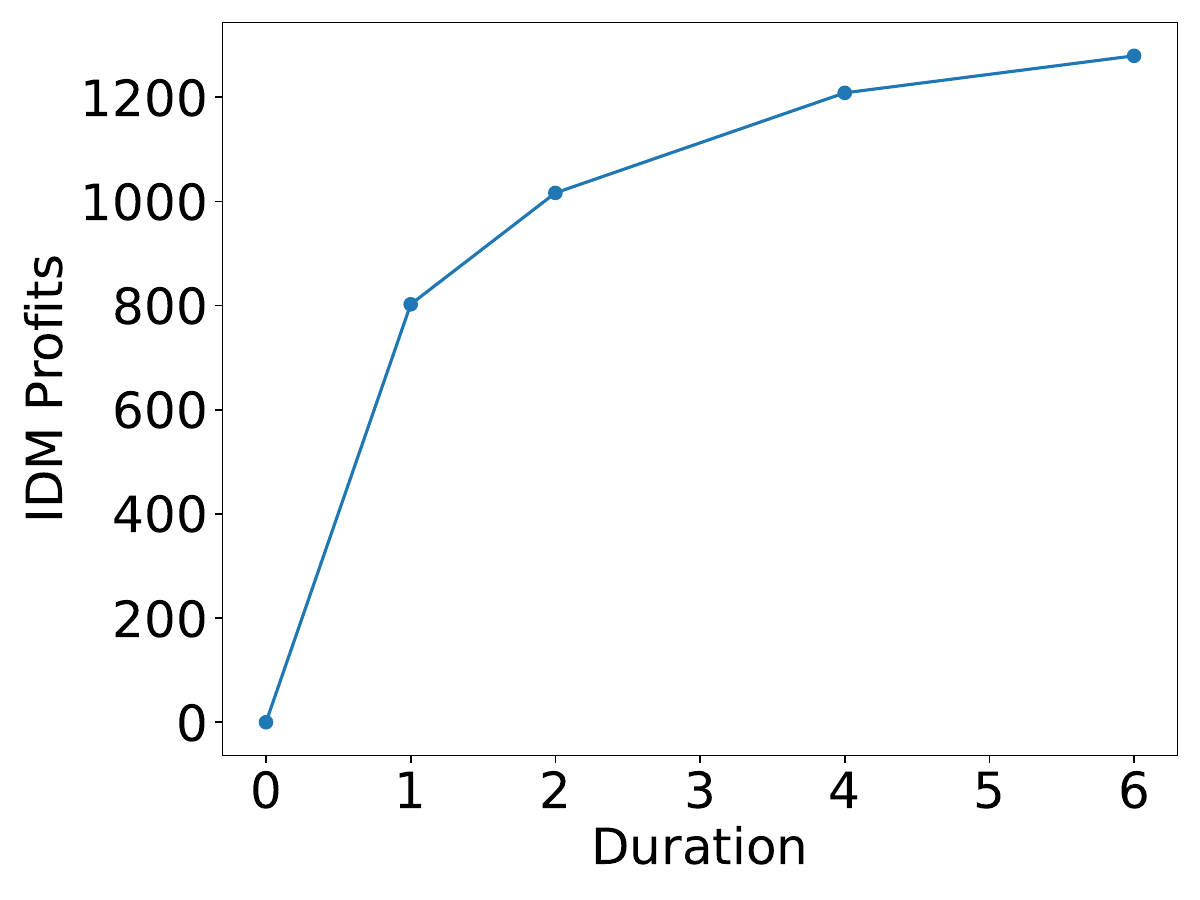}
    \caption{\label{fig:fcr_power_experiment} The left panel shows the relationship between the bid size in the FCR and the duration ratio, while the right panel depicts the how the capacity to power ratio affects the profits in the IDM.}
\end{figure}

In reality, price patterns fall between the two extremes depicted in Figure \ref{fig:price_patterns}, showing a more nuanced effect of duration on profits. We explore this relationship in Figure \ref{fig:fcr_power_experiment} by computing the profits of a 10MW storage with varying energy capacity resulting in different storage duration. The experiment is carried out with price data from January 2023. In the experiment, the energy capacity varies from $10$ MWh, that is, a duration of $1$, to $60$ MWh, i.e., a duration of $6$. 

The left panel of Figure \ref{fig:fcr_power_experiment} shows how the duration of the storage on the IDM increases with the size of the FCR bid and the right panel shows the connection between duration and overall average profits for one day of trading with the rolling intrinsic. This analysis reveals that there is a concave and increasing relationship between duration and per-MW revenue on the IDM.

Overall, this demonstrates that due to the way power and capacity are regulated in the FCR market, revenues in the IDM market fall more slowly than may naively be expected when increasing the battery power committed in the FCR market. Since FCR revenues per MW are constant in the FCR bid size, it also shows that strategies with a higher power committed to FCR tend to be more attractive for the 10MW/10MWh battery under consideration. 

\subsection{Pre-selection of FCR Strategies}
\begin{table}[t]
    \centering
    \begin{small}
        \begin{tabular}{lllllllll}
            \toprule
                        & \multicolumn{4}{c}{{\bf 2023}} & \multicolumn{4}{c}{{\bf 2024}} \\
            \cmidrule(rl){2-5} \cmidrule(rl){6-9}
            {\bf Strategy} & {\bf Overall} & {\bf IDM} & {\bf FCR} & {\bf Best} & {\bf Overall} & {\bf IDM} & {\bf FCR} & {\bf Best} \\
            \midrule
                (5,5,5,8,8,8) & 2815 & 827 & 1988 & 14 & 3559 & 750 & 2809 & \textbf{19} \\
                (8,8,8,5,5,5) & 3079 & 1077 & 2003 & 3 & 3535 & 919 & 2616 & 0 \\
                (8,8,8,0,0,0) & 2992 & 1744 & 1247 & 23 & 2945 & 1533 & 1412 & 3 \\
                (8,8,8,0,0,5) & 3033 & 1601 & 1432 & 20 & 3057 & 1399 & 1658 & 4 \\
                (8,8,8,0,0,8) & 2985 & 1442 & 1543 & 6 & 3061 & 1256 & 1805 & 2 \\
                (8,8,8,0,5,0) & 2928 & 1436 & 1492 & 0 & 3029 & 1253 & 1777 & 0 \\
                (8,8,8,0,5,5) & 2969 & 1292 & 1677 & 1 & 3124 & 1102 & 2022 & 1 \\
                (8,8,8,0,5,8) & 2899 & 1111 & 1788 & 3 & 3128 & 959 & 2169 & 1 \\
                (8,8,8,0,8,0) & 2863 & 1224 & 1639 & 3 & 3046 & 1051 & 1995 & 1 \\
                (8,8,8,0,8,5) & 2874 & 1051 & 1824 & 2 & 3136& 896 & 2241 & 0 \\
                (8,8,8,0,8,8) & 2799 & 865 & 1935 & 2 & 3143 & 755 & 2388 & 0 \\
                (8,8,8,5,0,0) & \textbf{3174} & 1601 & 1573 & \textbf{26} & 3388 & 1381 & 2007 & 2 \\
                (8,8,8,5,0,5) & \textbf{3185} & 1427 & 1758 & 17 & 3478 & 1226 & 2252 & 1 \\
                (8,8,8,5,0,8) & 3110 & 1241 & 1869 & 3 & 3476 & 1077 & 2399 & 3 \\
                (8,8,8,5,5,0) & 3090 & 1273 & 1818 & 4 & 3451 & 1080 & 2371 & 1 \\
                (8,8,8,5,5,8) & 2998 & 885 & 2114 & 0 & 3540 & 777 & 2763 & 2 \\
                (8,8,8,5,8,0) & 2989 & 1024 & 1965 & 1 & 3460 & 871 & 2589 & 1 \\
                (8,8,8,5,8,5) & 2973 & 823 & 2149 & 4 & 3549 & 714 & 2835 & 1 \\
                (8,8,8,5,8,8) & 2894 & 634 & 2260 & 3 & 3554 & 572 & 2982 & 1 \\
                (8,8,8,8,0,0) & \textbf{3242} & 1473 & 1769 & \textbf{77} & 3604 & 1241 & 2363 & \textbf{29} \\
                (8,8,8,8,0,5) & \textbf{3221} & 1267 & 1953 & \textbf{27} & \textbf{3683} & 1075& 2608 & 11 \\
                (8,8,8,8,0,8) & \textbf{3141} & 1076 & 2064 & 20 & 3682 & 927 & 2756 & \textbf{17} \\
                (8,8,8,8,5,0) & 3127 & 1114 & 2013 & 10 & 3663 & 936 & 2727 & 8 \\
                (8,8,8,8,5,5) & 3106 & 908 & 2198 & 11 & \textbf{3745} & 773 & 2973 & 7 \\
                (8,8,8,8,5,8) & 3025 & 716 & 2309 & 2 & \textbf{3750} & 630 & 3120 & 11 \\
                (8,8,8,8,8,0) & 3014 & 854 & 2160 & 9 & 3664 & 718 & 2946 & 12 \\
                (8,8,8,8,8,5) & 2999 & 654 & 2345 & \textbf{28} & \textbf{3751} & 560 & 3191 & \textbf{20} \\
                (8,8,8,8,8,8) & 2925 & 469 & 2456 & \textbf{46} & \textbf{3761} & 422 & 3338 & \textbf{116} \\
            \bottomrule
        \end{tabular}
    \end{small}
    \caption{\label{tab:strategies} Results of static strategies for 2023 and the first three quarters of 2024. The reported profits are in EUR per day and the results of the column best indicate on how many days a strategy is the best strategy. Boldface in the columns \emph{Overall} and \emph{Best} indicate the best five strategies in the respective year with regard to profits and number of times a strategy is the best strategy.}
\end{table}

In our case, there are in total $9^6$ different possible FCR bids, making it impractical to backtest all of them. The findings in the last section and initial experiments show that allocation of a large proportion of the BESS' capacity to the FCR market is preferable. However, the afternoon and evening EFA blocks four, five, and six can on some days exhibit substantial volatility with big price spikes in the intraday market. In these instances having more power capacity to trade in the IDM market to benefit fully from these spikes may outweigh the lost revenue in the FCR market.

Based on these findings, we choose the $28$ strategies listed in Table \ref{tab:strategies} as the set $\Mc$ for backtesting. All strategies are computed by reoptimizing once every minute in the rolling intrinsic strategy with a snapshot of the order book that contains the best four orders in either direction. In order to separate days and get revenues per day of trading, we start trading at 7 p.m. of the previous day for every trading day and assume an initial storage level of 2 MWh, which proved to be a good choice in our tests.

Table \ref{tab:strategies} reveals that the best strategy in the year $2023$ was (8,8,8,8,0,0), reducing the allocation of power to FCR in the late afternoon and evening blocks to take advantage of price spikes in the IDM in the corresponding hours. Generally, strategies that sell less power in the FCR market in these blocks perform best in $2023$. Interestingly in $2024$ the situation changes and the strategy (8,8,8,8,8) that commits the maximum capacity on the FCR market comes out on top with respect to average daily profits and the number of days the strategy dominates all the other strategies.

Looking at the IDM and FCR profits, we find average values of EUR $1111$ and EUR $1905$ in $2023$ and EUR $959$ and EUR $2468$ in $2024$. This implies that while the IDM market appears to have become less profitable for storage in $2024$ the opposite appears to have happened to the FCR prices, which explains the change in the type of optimal strategies between years. Overall, the average profit is higher in $2024$.

We furthermore observe that there are strategies that are not particularly good on average but are the best strategy on a significant amount of days. Examples include strategies (8,8,8,8,8) and (8,8,8,8,5) in $2023$. This supports our approach of selecting $\Sc \subseteq \Mc$ strategies that complement each other in an optimal way, rather than simply choosing the strategies that perform best on average.

Although we limit the set $\Mc$ to a reduced strategy pool with only $28$ strategies, picking up the optimal strategy for any given day from the such pool, equivalent to a classification problem of $28$ classes, which is made difficult by the inherent class imbalance in the data. To this end, we propose to select a subset $\Sc\subseteq \Mc$ to reduce the size of the strategy pool while staying close to the revenue potential of the entire set $\Mc$. 

\begin{figure}[t]
    \centering
    \includegraphics[width=0.525\textwidth]{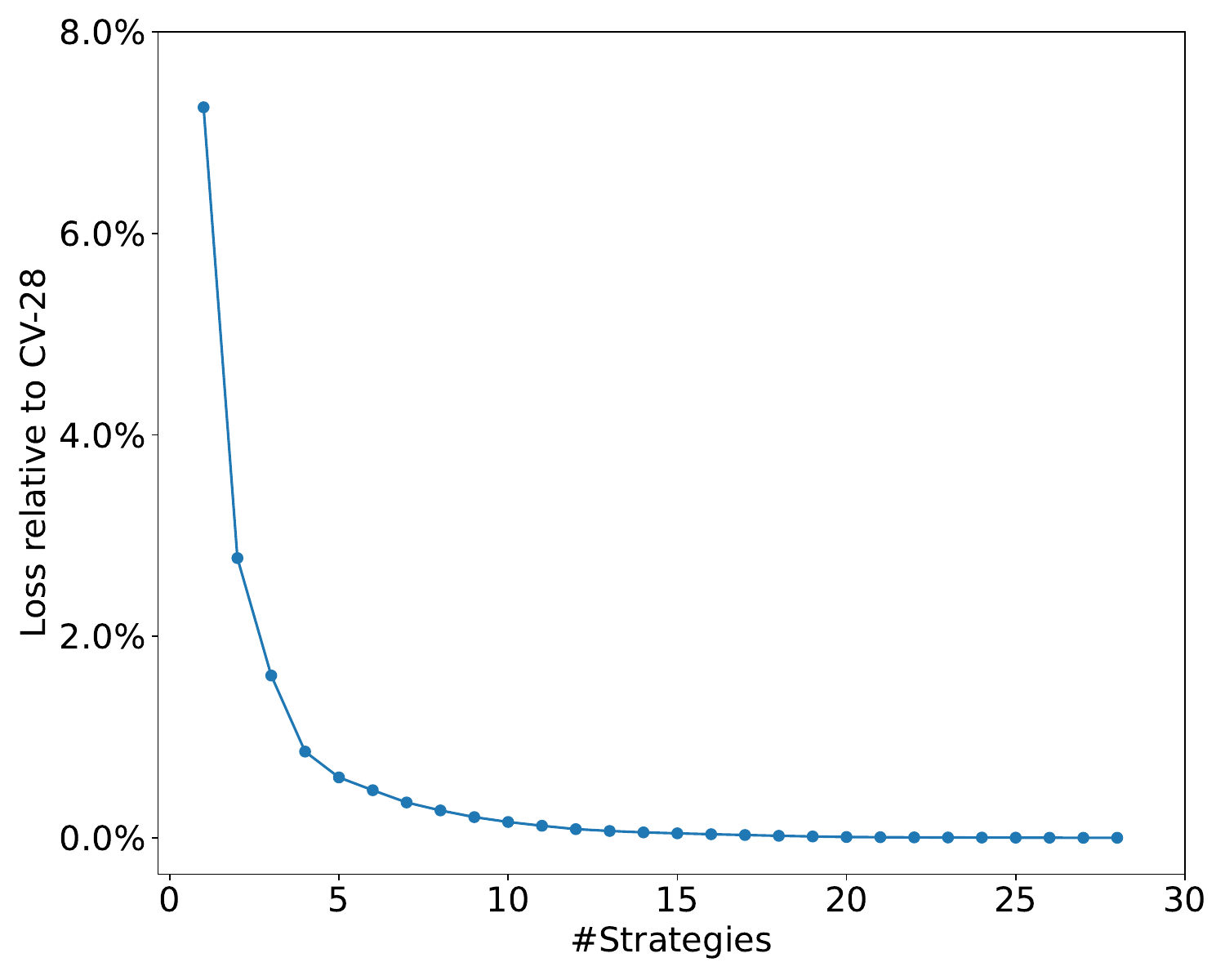}
    \caption{\label{fig:impr} Number of strategies $S$ versus profit percentages loss relative to CV-28. The results comes from 01/01/2023 to 30/09/2024}
\end{figure}

Figure \ref{fig:impr} shows the loss with optimal strategy selection relative to the clairvoyant strategy that always chooses the best strategy from $\Mc$ under the assumption that the best among the $S$ chosen strategies is used on every day. It can be clearly seen that for a small number of strategies $S$ the loss reduces rapidly but starts to level off around $S = 4$. This presents valuable insights in choosing the optimal number of strategies: Too many classes not only increase the difficulty of the task, but also offer only marginally increased profits.


\subsection{Features}
To select the most profitable FCR bidding strategy for a given day, we train our classification model on a set of features designed to capture the prevailing and expected market conditions. This feature set comprises primary predictors derived from market data and forecasts, as well as carefully selected interaction terms to model non-linear relationships between market drivers.

Our model is built upon a foundation of four groups of base features:
\begin{enumerate}
    \item Price of the day-ahead auction (DAA): Naturally, the day-ahead market prices are good predictors of continuous intraday prices \citep{unie19,ma20,nara20}. We incorporate day-ahead auction prices as input features, focusing on four selected regions — ‘DE-LU’, ‘IT-North’, ‘NO2’, and ‘SE4’ — due to their relatively low price correlations, which introduces greater regional diversity into the model. To better capture temporal market dynamics, we compute the average day-ahead prices within each EFA block. In addition, we calculate the standard deviation for each EFA block to reflect price volatility, which can be informative to capture risk and uncertainty in market behavior. In all this yields $4 \times 6 \times 2 = 48$ features for our model.
    
    \item VRES producers typically sell their products on the DAM on the basis of production forecasts. Larger forecast RES production therefore also yields to potentially larger absolute forecast errors, prompting more significant rebalancing on the intraday market. Hence, renewable generation forecasts are a major driver of intraday prices and, in particular, intraday price variance. This in turn influences the profitability of trading with the storage on the intraday market and is therefore a good feature in our strategy selection. See also \citep{karanfil2017role, goodarzi2019impact, kulakov2019impact, kuppelwieser2023intraday} for a discussion on the impact of renewable generation forecasts on intraday prices. Therefore, we also utilize public power forecasting (PPF) data, which includes forecasts for solar generation, onshore wind, offshore wind, and electricity load in Germany. To align with the temporal structure of our model, we calculate the average values of these forecasts within each EFA block, as well as the standard deviation of hourly values within each block, which yields an additional $6 \times 2 \times 4 = 48$ features.
    
    \item Historical FCR clearing prices serve as strong indicators of the potential profitability of the FCR market for the next day. To take advantage of this information, we include the FCR clearing price of the previous day for each of the $6$ EFA blocks as features in our model. This helps capture recent market conditions and short-term trends relevant to price formation.

    \item We add eight dummy features to capture temporal patterns. These include the \textit{weekday}, ranging from 0 (Sunday) to 6 (Saturday); a \textit{weekend indicator}, equal to 1 for Saturday and Sunday and 0 otherwise. Furthermore, we include the \textit{day of year}, denoting the $i$-th calendar day (1--365); and a \textit{trend variable}, representing the $i$-th day in the backtesting period (0--636). To account for seasonality, we further include both \textit{annual seasonality} terms\footnote{
$\sin\!\left( \tfrac{2\pi \cdot \text{day of year}}{365} \right)$,
$\cos\!\left( \tfrac{2\pi \cdot \text{day of year}}{365} \right)$}
as well as 
\textit{weekly seasonality} terms\footnote{
$\sin\!\left( \tfrac{2\pi \cdot \text{weekday}}{7} \right)$,
$\cos\!\left( \tfrac{2\pi \cdot \text{weekday}}{7} \right)$}. 
\end{enumerate}

While the base features are informative, the optimal strategy often depends on the complex interplay between different market drivers (e.g., the impact of a high wind forecast may be different on a day with high prior FCR prices versus low ones). To model these relationships, we introduce quadratic interaction terms. However, creating all possible pairwise interactions from the base features would result in an unmanageably large feature set (roughly $10000$ interactions), increasing the risk of overfitting and making the backtesting process computationally prohibitive. We therefore adopt a multi-step heuristic approach to construct a concise yet powerful set of interaction features:

\begin{enumerate}
    \item Focus on Primary Drivers: We generate interactions using only the average values of DAA prices and PPF forecasts for each EFA block, excluding the standard deviations to reduce the initial scope.

    \item Prioritize Cross-Group Interactions: We hypothesize that the most significant effects arise from the interplay between different feature groups. We therefore limit the interactions to two specific types: \emph{Day-ahead price averages} $\times$ \emph{Historical FCR prices} as well as \emph{Renewable forecast averages} $\times$ \emph{Historical FCR prices}

    \item Remove Redundancy via Correlation Filtering: This process results in an intermediate feature set of approximately 400 base and interaction terms. To mitigate multicollinearity and further reduce dimensionality, we perform a final filtering step. We compute the Pearson correlation matrix for this expanded set of features and iteratively remove one feature from any pair with a correlation coefficient exceeding a threshold of 0.94.
\end{enumerate}
This structured approach yields a final, manageable set of under 300 features that captures key linear and non-linear market dynamics.

\subsection{Rolling Horizon Evaluation} \label{sec:rolling_classifier}
\begin{figure}[t]
    \centering
    \includegraphics[width=0.5\columnwidth]{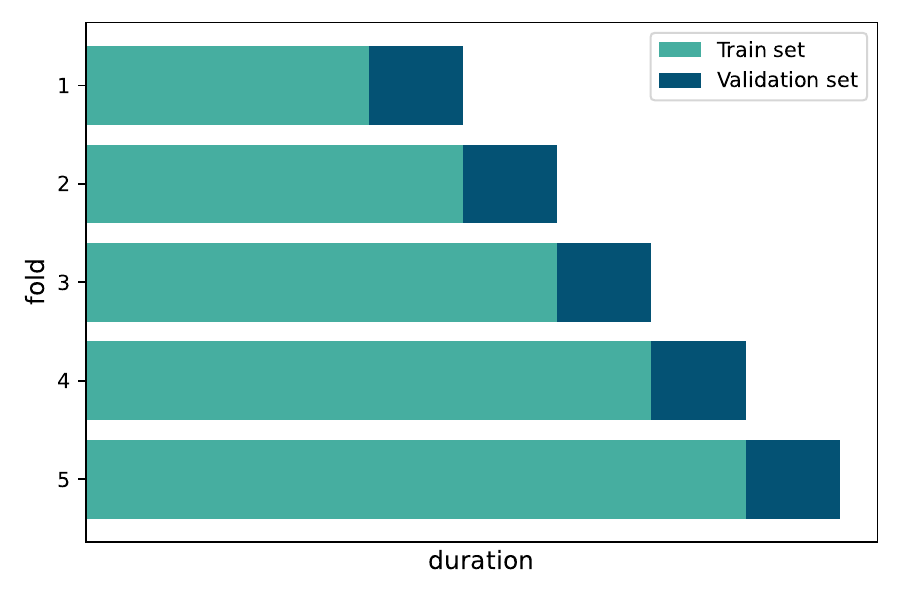}
    \caption{\label{fig:anch_val} Anchored walk-forward validation.}
\end{figure}

In our out-of-sample model evaluation, we mimic how the model would be used in trading practice on a day-to-day basis. For any given day in the out-of-sample window, we estimate an optimal policy to bid on the FCR market for one day with a model trained with data from the last $N$ days. Then we implement the policy and record its profits. Subsequently, we refit the model for the next trading day updating the training data with the last day. We generate out-of-sample profits by rolling through our entire data in this fashion.

More specifically, we evaluate the out-of-sample performance of the model in the rolling horizon fashion described above using data from January 1, 2023, to September 30, 2024. Each day, a training window of $N=240$ days is used. In this way, we generate a set of 396 out-of-sample daily profits covering the period from August 30, 2023, to September 30, 2024.\footnote{Note that since we use lagged prices, we actually need $241$ days insample data to generate data for $240$ training days.} Evaluating performance on this large and diverse set of days helps ensure that the results are not unduly influenced by any specific time frame or by particular patterns or anomalies that might appear in a smaller subset of the data.

Every time we retrain the model for a day of trading, we first solve equation \eqref{eq:best_group_selector} to select a subset of size $S = 3$ from $\Mc$ that achieves the highest average profits on the training sample, and then train an XGBoost classifier to further identify the optimal strategy from this subset. The size of $S=3$ strikes a good balance between the profit of the combined strategies and the difficulty of the ensuing classification problem (see Section \ref{ssec:FCR_strategies}).

When using XGBoost, we optimize the hyperparameters of the method. Specifically, we optimize the parameters \emph{eta} ($0.01$, $0.05$, $0.1$), \emph{gamma} ($0$, $0.5$, $1$, $2$), \emph{subsample} ($0.8$, $1$), \emph{colsample\_bytree} ($0.8$, $1$), \emph{max\_depth} ($3$, $4$, $5$), \emph{n\_estimators} ($200$, $400$) using cross validation. We use random grid search to find the best combination of hyperparameters selecting from the values in brackets. In particular, we randomly chose $20$ possible combinations of hyperparameters without replacement and test their performance in the validation sets and additionally include the best parameter combination found for the last day in the sample. 

Due to the periodicity and autocorrelation inherent in sequential data, traditional cross-validation may disrupt temporal dependencies, resulting in a significant discrepancy between the validation and test sets. We use anchored walk-forward validation, which preserves the sequential order by always validating on future data, to find the best set of hyperparameters, see Figure \ref{fig:anch_val}. Specifically, we employ five folds, with the last 15 days in each fold reserved for validation. Note that by anchoring, the training data grows in later folds. Instead of using accuracy or F-score as the criterion for hyperparameter tuning, we adopt use the negative trading profit in the validation sets as loss function for hyperparameter tuning. This is because accuracy and F-score may overlook misclassifications that have a significant negative impact on profitability, whereas a profit-based score penalizes such errors more appropriately.

When generating and evaluating the generated policy for every day, we add up FCR and intraday revenues, the latter of which is the result of executing the RI on the order book data for that specific day. The total profits are calculated by subtracting the battery degradation cost as described in Section \ref{sec:setting} from these revenues. As for the training data, in order to run the rolling intrinsic, we use snapshots of the LOB consisting of the top four prices for each traded product and rerun the intrinsic every minute. For every day, we start trading at 7 p.m. on the previous day and assume a starting storage level of 20\% (2MWh). Furthermore, we use the historical market clearing price of the FCR market to calculate the FCR revenues of the different strategies.

\subsection{Results \& Discussion}
\begin{table}[t]
    \centering
    \begin{tabular}{lrrrrrrr}
        \toprule
        Strategy & FCR & IDM & Overall  & \% of CV-28 & Equals CV-3 & Equals CV-28 & Beats LCS\\
        \midrule
        CV-28 & 1113 & 407 & 1520 & 0.0 & 46.4 & 100.0 & 59.3\\
        CV-3 & 1127 & 370 & 1497 & -1.5 & 100.0 & 46.4 & 27.8\\
        LCS & 1146 & 313 & 1459 & -4.0 & 72.2 & 38.9 & 0.0\\
        DB & 1063 & 349 & 1412 &  -7.1 & 36.9 & 20.1 & 16.4\\
        SB  & 969 & 433 & 1402 & -7.8 & 4.3 & 0.3 & 30.3\\ \addlinespace[5pt]
        Only FCR & 1238 & 170 &  1408 & -7.4 & 57.1 & 37.1 & 14.14 \\
        Only IDM & 0 & 774 &  774 & -49.1 & 0.0 & 0.0 & 17.7 \\
        \bottomrule
    \end{tabular}
    \caption{\label{tab:results} Overall out-of-sample profits (thousand Euro) of all strategies split into FCR and IDM profits. The last four columns give the shortfall from CV-28 (in \% of profit), the fraction of days (\%) where policies take the same bidding decisions as CV-3 and CV-28, as well as the fraction of days where policies outperform the LCS strategies, respectively.}
\end{table}

In this section, we compare the results of the trained classifier strategy (LCS) described in the previous sections with some benchmark strategies. The first benchmark is the 8-8-8-8-0-5 strategy, also referred to as the static baseline (SB), which is the most profitable single strategy in $\Mc$ calculated for the whole data. In addition, we compare our policy with the strategy that selects the strategy that was most profitable for the training data associated to a specific day. We refer to this as the dynamic baseline (DB). The perfect dynamic strategy denoted by CV-3 denotes an ideal policy that always selects the best performing strategy from the optimal group $\Sc$ on any given day.  Finally, as another theoretical upper bound, we include the CV-28 that always chooses the best strategy from $\Mc$ for every out-of-sample day. Furthermore, we also evaluate the strategies to bid only on the IDM and to bid the maximum amount allowed on the FCR market, that is, the strategies (8,8,8,8,8) and (0,0,0,0,0,0).

The results of this analysis are summarized in Table~\ref{tab:results}, which allows for the following conclusions: First, our proposed LCS performs exceptionally well, achieving an overall profit that is only $4.0$\% lower than the theoretical maximum achievable with perfect foresight (CV-28). This gap represents the combined loss from two distinct steps: the initial selection of the top-three strategy pool in \eqref{eq:best_group_selector} and the subsequent classification error. Critically, the gap shrinks to a mere $1.5$\% when compared to CV-3, the clairvoyant strategy restricted to the same three strategies. This demonstrates the high accuracy of the XGBoost classifier in selecting the best option from the available pool on any given day. In stark contrast, the naive dynamic (DB) and static (SB) benchmarks fall short of the theoretical maximum by $7.1$\% and $7.8$\%, respectively, highlighting the significant value added by the learning-based approach. Furthermore, strategies focusing exclusively on a single market are clearly suboptimal; the 'Only FCR' strategy underperforms the CV-28 benchmark by $7.4$\%, while the 'Only IDM' strategy lags by a substantial $49.1$\%. The later results also shows that the FCR market is more profitable than the IDM in the observation period.

Second, a breakdown of the revenue sources reveals that the performance difference between the top strategies is primarily driven by profits from the intraday market. Both CV-28 and CV-3 achieve higher IDM revenues than our LCS. This indicates that the main challenge, and the primary source of the remaining performance gap, lies in perfectly identifying the specific days where the intraday market offers exceptionally high profit opportunities.

Third, the effectiveness of the classifier is further confirmed by analyzing the daily decision alignment shown in Table~\ref{tab:results}. LCS selects the same strategy as CV-3 for an impressive 72.2\% of the days, a high hit rate that directly corresponds to its strong profit performance. The alignment with the best overall strategy (CV-28) is naturally lower at $38.9$\%, as the LCS is limited to strategies from $\Sc$. The fact that the CV-3 strategy itself only matches the CV-28 decision on $46.4$\% of days underscores the importance of the initial strategy pool selection and explains why a gap between our LCS and the theoretical maximum remains. 
    
Finally, the daily decision metrics in Table~\ref{tab:results} provide further insight into the performance of the model. The CV-3 strategy, which has perfect foresight over the same pool of three strategies available to our LCS, outperforms it on only $27.8$\% of the days. This implies a high classification accuracy for the XGBoost model, as it correctly identifies the best strategy within its pool more than $70$\% of the time. A nuanced observation arises when comparing LCS to the SB strategy. While SB outperforms LCS on a higher fraction of days (30.1\%). This occurs because SB can be the strategy outside the curated three-strategy pool used by LCS; however, the marginal profits gained on those days are negligible, which explains why SB’s overall profit remains substantially lower.

The data also allows to evaluate the quality of our initial strategy pool selection. The fact that the globally optimal strategy from all $28$ candidates was contained within our chosen pool of three on $46.4$\% of the days confirms that our selection process effectively captures the best-performing strategy in nearly half of all trading scenarios. Finally, the results reaffirm the market dynamics of the study period: the 'Only FCR' strategy proved to be the single best choice on a remarkable $37.1$\% of all days, underscoring the dominance of the FCR market and explaining why the top-performing strategies are heavily weighted towards FCR commitments.

\begin{figure}[t]
    \centering
    \includegraphics[width=0.8\textwidth]{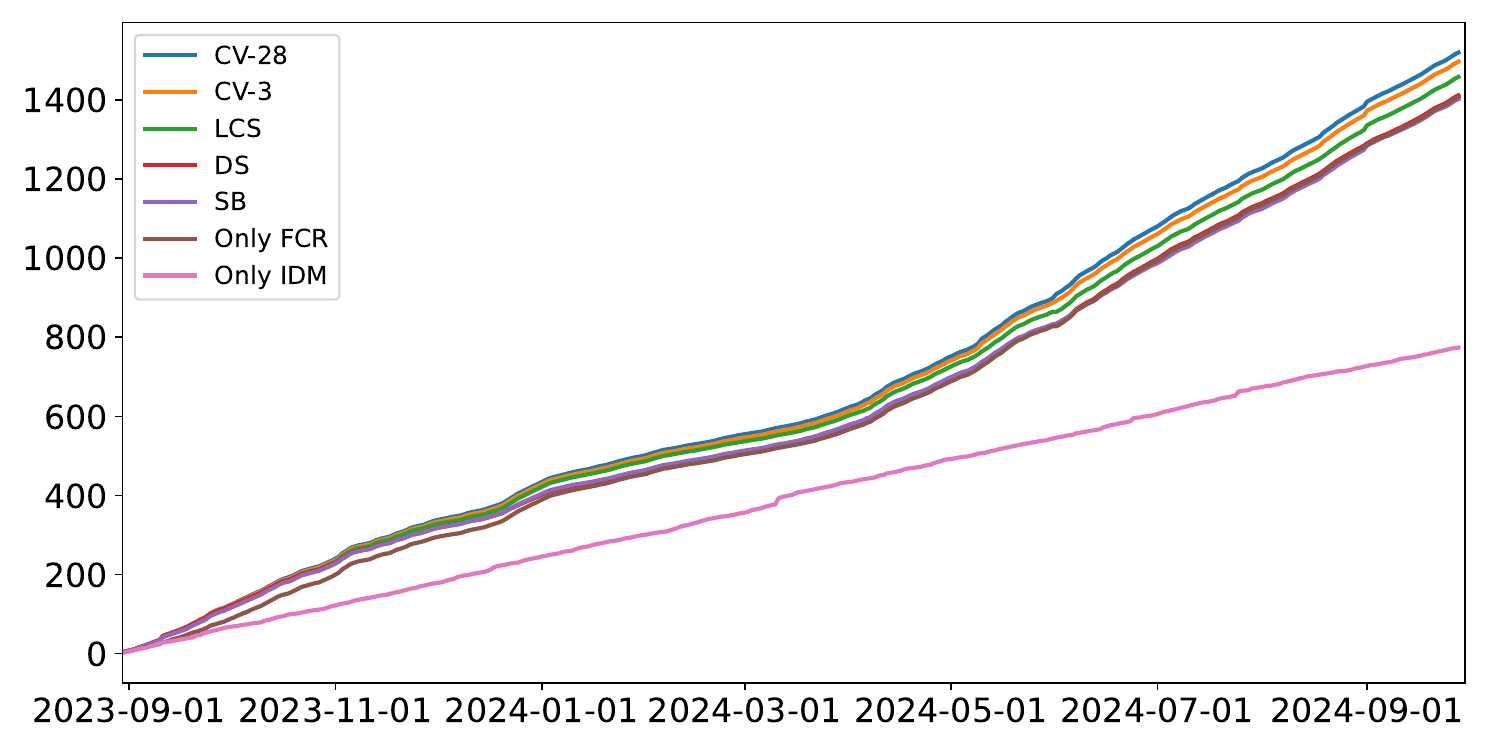}
    \caption{\label{fig:cumprofits} Cumulative profits for the LCS and all benchmarking strategies over the whole OOS period.}
\end{figure}
Figure \ref{fig:cumprofits} shows the cumulative profits of the strategies during the out-of-sample period. We can see that \emph{Only IDM} is clearly the worst strategy, while all mixed strategies earn profits in a comparable range. In accordance with the risk-free nature of the RI, the profits increase monotonically over time, and the slope substantially increases around the middle of the observation period, which is mainly due to the increase in profits in the FCR market.

\begin{figure}[t]
    \includegraphics[width = \columnwidth]{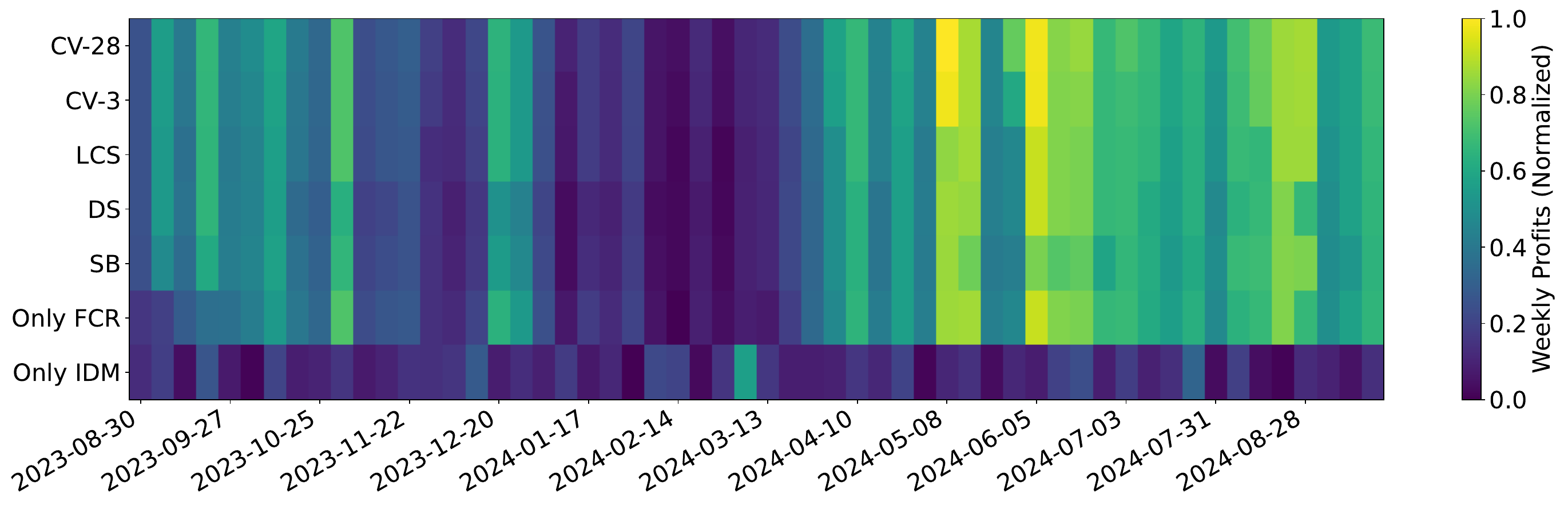}
    \caption{\label{fig:heatmap} A heatmap presents the normalized weekly profits across different strategies over the entire evaluation period. The horizontal axis denotes time (from August 2023 to August 2024), while the vertical axis lists the strategies. The color scale ranges from purple (low profits) to yellow (high profits).}
\end{figure}

Figure \ref{fig:heatmap} displays normalized weekly profits of different strategies \footnote{$\text{normalized weekly profits} = \frac{\text{weekly profits - min(weekly profits)}}{\text{max(weekly profits) - min(weekly profits)}}$}. The results reveal clear temporal and strategic patterns. In particular, the dynamic strategies (CV-28, CV-3, LCS) consistently achieve higher profits compared to the other four strategies. In particular, there is a pronounced increase in profitability between May and July 2024, when most dynamic strategies reach their peak performance (indicated by the yellow regions). In contrast, during late 2023 and early 2024, profits remain relatively low across all strategies. Furthermore, LCS exhibits a more stable and resilient performance compared to DS and SB. In particular, during the low-profit months from November 2023 to February 2024, LCS sustains higher profit levels, avoiding the sharp declines seen in DS and SB. Similarly, in March and April 2024, LCS continues to outperform both strategies, maintaining moderate profitability where DS and SB weaken. Moreover, during the peak period between May and July 2024, LCS tracks closely with CV-28 and CV-3, showing consistently high profit levels (yellow regions), while DS and SB display greater variability. This highlights that LCS is not only more robust in low-profit phases but also competitive with the strongest combined strategies in high-profit periods.

The computational requirements of our approach can be divided into two phases: the initial backtest and the daily operational deployment. The initial backtest, which involves simulating the RI performance for each FCR strategy over the entire dataset, is computationally intensive and requires approximately 7 hours per strategy. However, this process is highly parallelizable, as simulations for strategies are independent of each other. We managed this workload by performing the evaluations on 48 AWS Batch instances, each equipped with 4 vCPUs and 24 GB of memory.

In a live trading environment, the daily computational workflow is highly efficient. The process begins by updating the training data, which involves running a one-day backtest for each of the 28 candidate strategies using market data from the most recently completed day. This simulation step takes approximately one minute. Immediately following the data update, the optimal strategy pool is re-selected and the XGBoost classifier is retrained to generate a bid for the upcoming day. The entire daily process, from the initial data update to the final prediction, is completed in under two minutes on a standard desktop computer. This rapid execution time confirms that our LCS framework is not computationally prohibitive and is well suited for practical deployment in real-world trading operations.

\section{Conclusion}  \label{sec:conclusion}
This paper introduces a novel hierarchical approach to optimize BESS participation in the frequency containment reserve and intraday markets. Our method effectively decouples the complex problem into two manageable stages: a high-frequency rolling intrinsic algorithm handles the continuous intraday trading and state of charge management, while an XGBoost-based classifier selects the optimal FCR commitment from a pre-screened pool of complementary strategies. Our out-of-sample backtest on historical market data validates the practical value of this approach; the proposed \emph{learned classifier strategy} substantially outperforms static and naive dynamic benchmarks, closing a significant portion of the gap to theoretical maximum profit.
    
Despite its strong performance, the framework has limitations that open clear avenues for future research. A primary constraint is the reliance on a small and curated set of FCR strategies. Prevailing market conditions have heavily favored the FCR market, causing the most profitable strategies to be highly similar, allocating maximum capacity to FCR. This clustering creates a difficult classification problem with severe class imbalance, limiting the performance ceiling of the LCS. Furthermore, for tractability, our model assumes zero-price bids in the pay-as-cleared FCR auction and uses the suboptimal myopic RI for intraday trading.

Future work should directly address these limitations. First, enhancing the intraday trading algorithm is a key priority. By moving beyond the RI to more sophisticated methods such as reinforcement learning, the profitability of the IDM could be increased. This would naturally lead to a more diverse set of viable strategies, mitigating the class imbalance and making the classification task more effective. Second, a promising extension is to endogenize the FCR bid price. Instead of assuming zero-price bids, a forecasting model could estimate the opportunity cost of FCR participation, effectively the IDM profit foregone for each EFA block. This would enable a more sophisticated bidding strategy that optimizes both the quantity and the price of the FCR commitment, further enhancing overall profitability.

\bibliographystyle{abbrvnat}
\bibliography{bibliography}  
\end{document}